Effect of phase separation induced supercooling on magnetotransport properties of epitaxial La$_{5/8-y}$Pr$_y$Ca$_{3/8}$MnO$_3$ (y≈0.4) thin film

Sandeep Singh,[1,2] Geetanjali Sharma,[1] P. K. Siwach,[1] Pawan Kumar Tyagi,[2] K. K. Maurya and H. K. Singh[1]*

[1]National Physical Laboratory (Council of Scientific and Industrial Research), Dr. K. S. Krishnan Marg, New Delhi-110012, India
[2]Department of Applied Physics, Delhi Technological University, Delhi-110042, India

Abstract

Thin films of La$_{5/8-y}$Pr$_y$Ca$_{3/8}$MnO$_3$ (y≈0.4) have been grown on single crystal SrTiO$_3$ (001) by RF sputtering. The structural and surface characterizations confirm the epitaxial nature of these film. However, the difference between the ω-scan of the (002) and (110) peaks and the presence of pits/holes in the step-terrace type surface morphology suggests high density of defect in these films. Pronounced hysteresis between the field cool cooled (FCC) and field cooled warming (FCW) magnetization measurements suggest towards the non-ergodic magnetic state. The origin of this nonergodicity could be traced to the magnetic liquid like state arising from the delicacy of the coexisting magnetic phases, viz., ferromagnetic and antiferromagnetic-charge ordered (FM/AFM-CO). The large difference between the insulator metal transitions during cooling and warming cycles (T$_{IM}^C$~64 K and T$_{IM}^W$~123 K) could be regarded as a manifestation of the nonergodicity leading to supercooling of the magnetic liquid while cooling. The nonergodicity and supercooling are weakened by the AFM-FM phase transition induced by an external magnetic field. T$_{IM}$ and small polaron activation energy corresponding the magnetic liquid state (cooling cycle) vary nonlinearly with the applied magnetic field but become linear in the crystalline solid state (warming cycle). The analysis of the low temperature resistivity data shows that electron-phonon interaction is drastically reduced by the applied magnetic field. The resistivity minimum in the lower temperature region of the self-field warming curve has been explained in terms of the Kondo like scattering in the magnetically inhomogeneous regime.

*Corresponding Author; Email: hks65@nplindia.org

**Introduction**

Phase separation (PS) is believed to be the key ingredient of the physics of doped rare earth manganites.[1-3] Extensive experimental and theoretical investigations spread over the last two decades have established as the most dominant mechanism on the composition-temperature (x-T) diagram of intermediate and low bandwidth manganites like $Nd_{1-x}Sr_xMnO_3$,[3,4] $Sm_{1-x}Sr_xMnO_3$[5,6] and $La_{1-x-y}Pr_yCa_xMnO_3$.[7-10] Amongst these materials, $La_{1-x-y}Pr_yCa_xMnO_3$ has emerged as the prototypical among the phase separated manganites. Different compositional and structural variants like bulk single crystal, polycrystals and thin films of $La_{1-x-y}Pr_yCa_xMnO_3$ have been investigated.[7-20] The pronounced nature of the PS has been established by the observation of (i) strong divergence of the zero filed cooled (ZFC) and field cooled warming (FCW) magnetization, (ii) pronounced hysteresis between the field cooled cool (FCC) and FCW magnetization, and (iii) prominent thermomagnetic hysteresis in the temperature and magnetic field (H) dependent resistivity ($\rho$) measured in cooling-warming cycles.[7-16] The coexistence of sub-micrometer scale ferromagnetic metallic (FMM) and antiferromagnetic/charge ordered insulator (AFM/COI) clusters has been demonstrated by a study by Uehara et al.[7] Their study has also shown that the AFM/COI phase appears explicitly in magnetotransport measurements only at $y \geq 0.3$. The coexisting FMM and AFM/COI clusters, directly impact the electrical transport by making it percolative, which is evidenced by huge residual resistivity ($\rho_0$) for $y \approx 0.4$ in the metallic regime.[7] The study of Ghivelder and Parisi[8] on bulk-polycrystalline $La_{5/8-y}Pr_yCa_{3/8}MnO_3$ ($y \approx 0.4$) has shown that COI phase appears at $T_{CO} \approx$ 230 K and subsequently undergoes transition to AFM and FM spin order at $T_N \approx$ 180 K and $T_C \approx$ 80 K, respectively. Large temporal relaxation in magnetization and resistivity has also been observed in this prototype PS system and has been attributed to the rapid spatial and temporal variations in the relative fraction of FMM and AFM phases.[8] Further, the theoretical study by these authors has predicted that interplay between temperature and separation of the system from equilibrium could create multiple blocked states.[8] Sharma et al.[9] studying a similar material have established the existence of a liquid like magnetic state in the phase separated regime, which transforms cooperatively to a randomly frozen glass like phase at low temperature. The frozen glass like phase (termed as strain glass) is believed to arise from the presence of martensitic accommodation strain.[9] Wu et al.[10] have demonstrated that in $La_{5/8-y}Pr_yCa_{3/8}MnO_3$ ($y \approx 0.4$) thin films the magnetic liquid like state exhibits a supercooled glass transition. This glass transition is believed to arise due to the presence of the accommodation strain caused by distinct structural symmetries of FMM and AFM/COI phases.[11] Their study has also provided evidence in favour of the non-ergodic nature of the magnetic liquid, which appears when the long range cooperative strain interactions hinder the cooperative dynamic freezing of the first-order AFM/COI–FMM transition.[9,10]

In thin films an additional degree of freedom to play with and tune the magnetic and transport properties becomes available in form of substrate induced strain. Due to the delicate nature of the phase separated state in manganites of smaller bandwidths like $Sm_{1-x}Sr_xMnO_3$[5,6] and $La_{1-x-y}Pr_yCa_xMnO_3$[12-18] the role of substrate induced strain becomes more pronounced. At reduced bandwidth the COI phase having AFM spin configuration acquires prominence over an appreciable range of x and hence the boundaries between the electronic phases, viz. paramagnetic insulator (PMI), FMM, ferromagnetic insulator (FMI), and AFM-COI become diffuse.[6-8] Investigations of the impact of substrate induced strain in $La_{0.67-x}Pr_xCa_{0.33}MnO_3$ thin films by Wu et al.[12] suggest occurrence of metastable phase mixtures, with the volume fraction of the constituent FMM and COI phases being controlled by the substrate induced strain. It was observed that ultrathin films having higher strain showed stronger phase separation tendency, as evidenced by a large hysteresis and slow relaxation behavior in transport measurements. The study of thin films of composition $La_{5/8-0.3}Pr_{0.3}Ca_{3/8}MnO_3$ deposition by laser MBE on different substrates carried out by by D. Gillaspie et al.[13] suggests that that the PS is sensitive to the substrate induced strain and that the long-range COI state is strengthened by tensile strain and suppressed by compressive strain. Dhakal et al.[14] have carried out detailed study of 30 nm thin films of $(La_{1-y}Pr_y)_{0.67}Ca_{0.33}MnO_3$ ($y$=0.4, 0.5, and 0.6) grown on $NdGaO_3$ substrates. Their results demonstrate that the nature of the phase separated state is composition dependence and that a fluid phase separated (FPS) state, which appears to be similar to the strain-liquid state in bulk compounds, exists at intermediate temperatures.[8] This FPS state is transformed to a metallic state by external electric field. They also suggest that the substrate-induced strain is a function of temperature and that the temperature induced variation of the long-range strain interactions plays a dominant role in determining the properties of thin films of phase-separated manganites. A study by Ward et al.[15] on $La_{0.325}Pr_{0.3}Ca_{0.375}MnO_3$ thin films demonstrates that the transport behaviour could be drastically different in a wire with a width comparable to the mesoscopic phase-separated domains inherent in the material and the parent thin film out of which the wires are fabricated. They observed an additional and more robust insulator metal transition (IMT) in a dimensionally confined structure like a wire. The roles of strain and 'kinetic arrest' across the first order transitions of a dynamic phase separated system has been probed by Sathe et al.[16] in $La_{5/8-y}Pr_yCa_{3/8}MnO_3$ ($y = 0.45$) thin films grown on substrates having different lattice mismatches. Their study reveals that the kinetics of first order phase transition is arrested across an AFM-COI to FMM transition and the arrested state behaves as a magnetic glass. Similar to structural glasses, these magnetic glass-like phases show evidence of devitrification of the arrested metastable AFM-COI phase to the equilibrium FMM phase with isothermal increase of magnetic field and/or isomagnetic field warming. Sathe et al. have also pointed out that the glass-like state and large scale dynamic phase separation is independent of the nature of strain and hence the 'kinetic arrest' dominates over strain in shaping large scale phase separation. The conductive atomic force microscopy (cAFM) measurements on single crystalline $(La_{0.4}Pr_{0.6})_{0.67}Ca_{0.33}MnO_3$ thin by Singh et al.[17] reveal that the distribution of conductivity across the surface is nonuniform. Furthermore, in the cooling cycle small patches of conductivity nucleate in a nominally insulating film surface, while during warming linear regions of insulating material form in a nominally conducting film surface. A study by Mishra et al.[18] has shown that in a phase separated system like $La_{5/8-y}Pr_yCa_{3/8}MnO_3$ (y = 0.45) thin film, strong strain field inhomogeneities developed during the strain relaxation mechanism produce phase separation at larger length scales. They have concluded that extrinsic disorder acts in a similar way to quenched disorder. The absence of extrinsic disorder results in a robust insulating phase with small phase separation while huge extrinsic disorder causes phase separation at larger length scales and shows an IMT in the absence of magnetic field at variance with the bulk compound. In $(La_{0.4}Pr_{0.6})_{0.67}Ca_{0.33}MnO_3$ thin films electric-field induced anisotropic transport has been observed in the FPS state by Jeen and Biswas.[19] According to them the main driving force for the anisotropy is the collective rearrangement of the FMM phase under electric fields.

Recently, Singh et al.[20] have investigated the tunability of the IMT in terms of the variation in the relative fraction of the coexisting FMM and AFM/COI phases in $La_{5/8-y}Pr_yCa_{3/8}MnO_3$ (y≈0.4) thin films (~42 nm). This study has clearly demonstrated that the supercooling transition temperature is non-unique and strongly depends on the magneto-thermodynamic path through which the low temperature state is accessed. In contrast, the superheating transition temperature remains invariant of the thermal cycling. However, the detail investigation of the impact of supercooling/superheating on the electrical transport in varying magnetic field has not been carried out. In the present paper we report detailed investigation on the structure, microstructure and magnetotransport properties of ~ 42 nm thin $La_{5/8-y}Pr_yCa_{3/8}MnO_3$ (y≈0.4) thin films grown by RF magnetron sputtering on (001) oriented $SrTiO_3$ (STO) substrate. Our study reveals that the nature of the electrical transport in the PM regime remains unaffected by the thermal cycling. In contrast, in the lower temperature region different scattering mechanisms appear to acquire dominance during the cooling and warming cycles. The parameters characterizing electrical transport are found to be non-linear in the supercooled regime and approach linearity in the superheated regime.

**Experimental Details**

Target (2″ diameter) of $La_{5/8-y}Pr_yCa_{3/8}MnO_3$ (y≈0.4) (LPCMO) was prepared by solid state method with desired amount of high purity $La_2O_3$, $Pr_2O_3$, $CaCO_3$, and $MnO_2$ compounds. The material was thoroughly mixed and heated at 900 °C for 24 hrs and further at 1000 °C with intermediate grinding. Then fine powder was pressed in the form of disc of 2″ diameter and ~3 mm thickness and sintered at 1200 °C for 24 hrs. The thin films were grown by RF magnetron sputtering in 200 mtorr of Ar + $O_2$ (80 % + 20 %) mixture on single crystal STO (001) substrate ($5x3x0.5$ $mm^3$) maintained at temperature ~ 800 °C. In the present study the lattice mismatch between the target used to deposie the films and the STO and substrate is ε ≈ -1.93 % [ε = ($a_t−a_s$) x100⁄ $a_s$ where $a_t$ and $a_s$ are the lattice parameters of the bulk target and substrate, respectively.] and hence the strain is tensile. In order to achieve optimum oxygen content the films were annealed at ~ 900 °C for 10 hr. in flowing oxygen. The film thickness was estimated from the X-ray reflectivity (XRR) measurements. The strcutructural and microstructural characteristics were probed by high resolution X-ray diffraction (HRXRD, PANalytical PRO X'PERT MRD, Cu-Kα1 radiation λ = 1.5406 Å) and atomic force microscopy (AFM, VEECO Nanoscope V), respectively. The temperature and magnetic field dependent magnetic and magnetotransport properties were measured by commercial MPMS and PPMS (both Quantum Design).

**Results and Discussion**

The experimental and simulated XRR curves are plotted in Fig. 1. The simulated XRR curve yield film thickness ≈ 42 nm. The structural information was extracted from 2θ-ω scan plotted in Fig. 2. Appearance of out of plane diffraction maxima (00ℓ) only in the 2θ-ω scan confirms highly oriented nature of the film. Out of plane lattice constant estimated from 2θ-ω scan is found to be $a_{cf}$ ≈ 0.3832 nm. This value is slightly smaller than average out of plane lattice constant ($a_{cb}$ ≈ 0.3842 nm) of the bulk used for the sputter deposition of the present film. The observed decrease in the out of plane lattice constant is due to the in-plane tensile due to the larger in-plane lattice constant of substrate (STO, $a_s$ ≈ 0.3905 nm) as compared to the bulk ($a_b$ ≈ 0.38426 nm). This tensile strain stretches the $MnO_6$ octahedron along the plane of the substrate and leads to the decrease in the out-of-plane lattice constant. The obvious consequence of the tensile strain is the strengthening of Jahn-Teller distortion and the superexchange interaction which favours AFM/COI phase. However, prolonged oxygen annealing carried out in the present study is expected to relax the strain to a great extent. This is supported by the fact there is only moderate difference between the out-of-plane lattice constant of bulk and film. To evaluate the in-plane growth nature of the film φ scans were measured. The φ scans of (001) plane of STO and LPCMO are plotted in inset of Fig. 2. The φ scans peak separation of 90° in STO as well as LPCMO confirm the four fold symmetry of the film and cube on cube coherent growth.

In order to acquire qualitative idea about the degree of defects and hence to probe the structural quality of the film, in-plane and out-of-plane rocking curves (ω scan) were measured. As shown in Fig. 3 the out-of-plane rocking curve of (002) reflection) has full width at half maximum (FWHM) $\Delta\omega \approx 0.63°$, which is higher than generally observed values for manganite films deposited by magnetron sputtering.[21] The in-plane rocking curve of (110) reflection has FWHM $\Delta\omega \approx 0.99°$, which is considerably larger than the out-of-plane value. The symmetry of the two rocking curves suggests that the film is nearly strain free. The long duration oxygen annealing and larger mismatch between the substrate (a = 0.3905 nm) and bulk (a= 0.3842 nm) in plane lattice parameters could be regarded as the major influences causing relaxation of strain.

The broadening of the rocking curve is generally attributed to the presence of (i) strain, (ii) dislocation density, (iii) mosaic spread, and (iv) curvature. As pointed out above the rocking curve broadening due to strain is expected to be very small. Although small variation in the FWHM of the rocking curves was observed as a function of the beam size, no linear dependence could be established between them. This rules out the contribution from curvature induced broadening. Thus the rocking curve broadening in the present case is attributed mainly to the mosaic spread and dislocation density. Since the FWHM variation with the beam size was not appreciable we believe that the dominant contribution to the peak broadening comes from the presence of dislocation arrays/network. The large difference between the FWHMs of (002) and (110) rocking curves suggests that the density of defects and mosaicity are different along the different planes, that is, the density distribution is anisotropic. In this regard it appears that the substrate film interface could have higher density of dislocations as compared to the epitaxial layers above. In fact it is well known that the dislocation networks present at the interface provide relaxation of substrate induced strain. The dislocation in the upper film layers are created due to the non-equilibrium energetic conditions that exist during the growth.

The epitaxial nature of the film is also reflected in the surface topography. A representative AFM image of the annealed film shown in Fig. 4, further confirms the epitaxial growth. The inset shows the AFM image of as grown film which is granular in nature with average roughness $\approx 1.41$ nm. Post deposition annealing in oxygen environment results in layer by layer growth/step terrace growth. This is clearly seen in the surface AFM image. Although the steps and terraces are not well defined, the average surface roughness is reduced to less than $\approx 0.47$ nm (nearly one unit cell). The appearance of holes and regular discontinuities in the individual layers could be regarded as an evidence of the lattice defects. These defects are generated due to the relaxation of the large strain between the substrate and the material during the oxygen annealing.

The temperature dependent magnetization M (T) was measured using ZFC, FCC and FCW protocols. The detailed analysis of the M (T) data has already been presented in reference 20. For the sake of convenience and continuity the main results are summarized here. The FM transition temperature ($T_C$) occurs at $T_C \approx 117$ K, $T_C^C \approx 63$ K and $T_C^W \approx 120$ K in the ZFC, FCC and FCW protocols, respectively. The protocol dependence of FM transition could be regarded as evidence of non-ergodicity. A significant difference between the $T_C$ in the FCC and FCW protocols coupled with the huge hysteresis between these two M (T) curves is a consequence of supercooling of the liquid like state due to the magnetic frustration caused by competing FM and AFM/COI interactions.[7-15] The explicit absence of the COI and AFM transitions could be attributed to defect induced quenching of AFM/COI in thin film form.[18,21] This has been explained in terms of the accommodation strain arising due to the distinct structural symmetry of the coexisting FM (pseudocubic) and AFM/COI (orthorhombic) phases[10,11,22]. The accommodation strain and the magnetic frustration would create multiple minima in the energy landscape of the system and hence could hinder the nucleation of the equilibrium low temperature state. Such a scenario in turn could give rise to a liquid like magnetic phase.[9,10] Further, the prominent divergence of the ZFC-FCW M (T) observed invariably in phase separated manganites as a signature of cluster glass state and originates due to the coexistence of FMM and AFM/COI phases below the Neel temperature $T_N$.[9-11] The cluster glass state is more common in intermediate bandwidth manganites having x $\approx$ ½[3,21,23,24] or over a much wider range of x in low bandwidth manganites like $Nd_{1-x}Ca_xMnO_3$,[25-27] and $Sm_{1-x}Sr_xMnO_3$[3,28-30]. The sharp drop in the ZFC curve in conjunction with the lower temperature reversibility of the FCC-FCW M (T) has been considered as a signature of cluster freezing.[8-10] In the reversible regime nearly temperature independent M (T) shows that the magnetic clusters are completely frozen or blocked and when the temperature is raised unblocking of the magnetic clusters occurs. The maximum in the FCW M (T) curve is the temperature where the clusters are completely unblocked.

The magnetic properties were further elucidated by measuring the magnetic field dependence of magnetization (M (H)) at several temperatures. The M (H) data measured at 10 K and 50 K is plotted in Fig. 5. At both temperatures virgin cycle M (H) show nearly identical field dependence and rise sharply up to ~2 kOe (the lower inset in Fig. 5). At H > 2 kOe the M (H) at 10 K bends and then increases linearly up to the highest applied magnetic field (H=50 kOe). In contrast, the M (H) measured at 50 K continues to increase beyond H=2 kOe, albeit with changing slope and then appears to saturate at H>30 kOe. The observed difference in the two virgin M (H) curves could be attributed to the different nature of the magnetic state. At 10 K the magnetic clusters are randomly frozen into a glassy state, while at 50 K is just above the temperature at which the clusters are unblocked completely (peak in the M-T curve). Hence at T=50 K, the FMM phase is dominant with possibility of AFM/COI droplets embedded into it. It is precisely due to this reason that M (H) at 50 K saturates. Beyond the virgin cycle the M (H) curve traces a typical loop at both the temperatures. In fact such behaviour of M (H) curves has been generally attributed to the metamagnetic nature of the magnetic ground state. In the LPCMO films under study, as explained in reference 20 and summarized above a metamagnetic glassy state is created by the competing FM and AFM/COI phases. The coercivity is observed to decrease from $H_C \approx \pm 737$ Oe at 10 K to $\pm 502$ Oe at 50 K (upper inset in Fig. 5). The remanent magnetization decreases from $M_r \pm 550$ emu/cm$^3$ at 10 K to $\pm 440$ emu/cm$^3$ at 50 K. The M (H) curves shows saturation tendencies at much higher field, around H~10 kOe in both cases.

The temperature dependent resistivity ($\rho$-T) measured at different values of H was measured in cooling and warming cycles (Fig. 6). In the cooling cycle $\rho$-T shows insulating behavior as shown by about six orders of magnitude rise in resistivity between $300 - 65$ K and IMT is observed at $T^C_{IM} \approx 64$ K. As T is lowered further down the $\rho$-T curve appears to saturate. In the warming cycle the $\rho$-T curve remains reversible with the cooling cycle $\rho$ (T) up to $T_g \approx 15$ K. As T is increased further up $\rho$ (T) decreases, approaching a minimum at $T_M \approx 45$ K (inset of Fig. 6). In the warming cycle the IMT shifts to a higher temperature and appears at $T^W_{IM} \approx 123$ K. In the PMI regime $\rho$-T curves overlaps with the one in cooling cycle. The two distinct transitions at $T_{IM}{}^C$ and $T_{IM}{}^W$ in cooling and warming cycles are separated by $\Delta T_{IM} \approx 59$ K. The observed thermal hysteresis in $\rho$ (T) is attributed to supercooling and superheating of the magnetic liquid consisting of FMM and AFM-COI phases and is an evidence of a first order phase transition.[20,23] The sharp drop in the $\rho$-T during the cooling cycle could be regarded as manifestation of dynamical magnetic liquid behavior. The saturation and reversible behavior of $\rho$-T at $T<T_g$ is due to the crossover from the liquid like state to the strain glass regime.[8-10,23] During the warming cycle the $\rho$-T minimum at $T_M \approx 45$ K has been regarded as a consequence of thermal devitrification of disordered SRG in to an ordered FMM.[16,18] Application of magnetic field enhances $T_{IM}{}^C/T_{IM}{}^W$, reduces the $\rho$–T hysteresis, and dilutes the first order nature of the IMT. Another noticeable effect of magnetic field is the disappearance of the $\rho$-T minimum at H=10 kOe. The observed $\rho$–T hysteresis and the minimum are related to the magnetic liquid and the SRG, respectively. The application of magnetic field increases the size and fraction of the FMM clusters and simultaneously reduces the size and fraction of the AFM/COI. This magnetic field induced increase in the FMM fraction is expected to (i) reduce the magnetic frustration, (ii) weaken the liquid like behavior and (iii) push the SRG towards a FMM. The variation of the $T_{IM}{}^C$ and $T_{IM}{}^W$ with magnetic is field plotted in Fig. 7 and the variation of $\Delta T_{IM}$ with magnetic field is plotted in the inset. The $T_{IM}{}^C$ shows a nonlinear dependence on H, while $T_{IM}{}^W$ increases linearly with it. The nonlinearity in the $T_{IM}{}^C$ – H could be regarded as a consequence of the liquid like behavior of the FMM-AFM/COI two phase mixture in the cooling cycle. In the warming cycle the FMM-AFM/COI competition is appreciably weakened and the materials behaves like nearly equilibrium FMM system. This results in linear $T_{IM}{}^W$ – H behaviour. Like the M (T) measurements, the cooling and warming cycle $\rho$-T data do not show any explicit evidence of AFM/COI, the absence of which could be attributed to the presence of lattice defects, such as dislocations. The presence of dislocations could destroy the AFM/COI order by weakening of the J-T distortion and delocalizing the charger carriers.[24]

Since the electron–lattice coupling through the J-T distortion of the $MnO_6$ octahedron favours an AFM order and in small bandwidth manganites it becomes strong so much so that it challenges the FMM phase and create magnetic frustration even at $T<T_C$. In the present case this is evidenced by the prominent hysteresis and the widely separated magnetic and electrical transition temperatures. Under the influence of a magnetic field the observed modifications in the electrical transport are generally associated with intrinsic nature of carriers and could be probe by the analysis of $\rho$-T data in the temperature range above $T_C$. In manganites the electron-lattice coupling can lead to the trapping of the charge carriers into a polaronic state, influencing the transport properties in the high temperature PM phase.[24,29,31-33] Hence, the temperature dependence of resistivity data in the PM regime is an important probe of the conduction mechanism. The $\rho$-T data at $T>T_C/T_{IM}$ have been analyzed in the framework of the Emin– Holstein[31] approach of small polaron hopping in the adiabatic limit given by the expression

$$\rho\,(T) = AT\exp\left(\frac{E_A}{k_BT}\right);\ A = \frac{2k_B}{3ne^2a^2\nu} \tag{1}$$

Here, 'n' is the polaron concentration, 'a' is the site-to-site hopping distance, '$\nu$' is the attempt frequency, and $k_B$ is the Boltzmann constant. $E_A$ is the activation energy, i.e., the height of the potential trap, and $E_A = E_P/2 - t$. In general, the overlap integral, 't' is so small that it could be neglected and then, $E_A \approx E_P/2$, or the polaron binding energy $E_P \approx 2E_A$.

The small polaron activation energy ($E_A$) was calculated from the fitting parameter derived from the best fit (data not plotted here) to eq. (1) in the temperature range $T_{IM}{\leq}T{\leq}300$ K. Variation of $E_A$ with magnetic field is plotted in Fig. 8(a). In absence of external magnetic field (H=0 kOe) the estimated value of $E_A$ is $\approx138$ meV. Such high values of the activation energy are typical of the low bandwidth manganite films[29] and are significantly larger than that of the large and intermediate bandwidth compounds like $La_{1-x}Sr_xMnO_3$[33], $Nd_{1-x}Sr_xMnO_3$.[24] Such high value of $E_A$ clearly suggests strong electron-lattice coupling through the J-T distortion. $E_A{\approx}138$ meV corresponds to $T_{IM}^{C}/T_{IM}^{W} \approx 64$ K/123 K in cooling and warming cycles. As the magnetic field is increased the value of $E_A$ decreases nonlinearly and at H=50 kOe, $E_A{\approx}112$ meV corresponds to $T_{IM}^{C}/T_{IM}^{W} \approx 184$ K/185 K in cooling and warming cycles. The variation of $T_{IM}^{C}/T_{IM}^{W}$ with $E_A$ is plotted in Fig. 8(b). The sharp rise in $T_{IM}^{C}$ at smaller magnetic field, that is, higher values of $E_A$ is a signature of the liquid like behaviour of the two phase mixture in the cooling cycle. This liquid like behaviour is not retraced in the warming cycle and hence $T_{IM}^{W}$ - $E_A$ behaviour is nearly linear.

Deep in the FMM regime, the $\rho$-T is characterized by (i) sharp decrease in the cooling cycle, (ii) appearance of a minimum at $T_M$ at H=0 kOe in the warming cycle which vanishes at H$\geq$10 kOe, (iii) relatively slower increase in the warming cycle. The low temperature $\rho$-T behaviour is not well studied in strongly phase separated manganites like the present compound. In order to explain the low temperature $\rho$-T behaviour in manganites several mechanisms have been in taken into consideration. The first is the Kondo like scattering given by, $\rho_k(T)\propto\ell nT$.[35] The second mechanism considered is an elastic scattering correction term of the type $\rho_c\propto T^{0.5}$, which accounts for the disorder enhanced strong Coulomb interaction between the carriers.

Such a term is generally observed in a disordered metallic system and it changes sign as a function of disorder in the system.[36] The third term is the electron-electron (e-e) scattering described by a contribution of the type $\rho_{ee} \propto T^2$.[33,36] The fourth contribution is the electron-phonon (e-ph) scattering, which at low temperatures acquires the form $\rho_{eph} \propto T^5$ and in view of the strong electron-lattice coupling through the J-T distortion is expected to be make significant contribution to the low temperature resistivity of low bandwidth manganites.[36]  In the present case we tried to fit the cooling and warming cycle low temperature $\rho$-T data measured at different magnetic fields using different combination of the above mentioned mechanisms. It is interesting to note that expect the zero magnetic field warming data, all the $\rho$-T curves at $T \ll T_{IM}$ were fitted very well by the following equation:

$$\rho(T) = \rho_0 + \rho_{ee}T^2 + \rho_{ph}T^5 \qquad (2)$$

Here, $\rho_0$ is the residual resistivity arising due to elastic scattering like electron impurity scattering. The $\rho$-T data of the zero field warming cycle could not be fitted with the equation 2. The appearance of the minimum in the H=0 warming cycle $\rho$-T prompted us to consider the Kondo like scattering. It has been shown that in inhomogeneous magnetic systems, e. g., those exhibiting glassy behaviour like the present films, Kondo like transport may appear in the low temperature regime.[35] The best fit to this was obtained by replacing the e-e scattering by the Kondo term ($\rho_k(T) \propto \ell n T$ ), that by the equation,

$$\rho(T) = \rho_0 - \rho_k \ell n T + \rho_{ph}T^5 \qquad (3)$$

However, the $\rho$-T data corresponding to the glassy state (T<18 K) could not be fitted even by this equation. The representative low temperature data along with the fitted ones are presented in Fig. 9. The fitting parameters $\rho_{ee}$ and $\rho_{ph}$ have been found to depend on the thermal cycling as well as the applied magnetic field. The values of both, $\rho_{ee}$ and $\rho_{ph}$ are found to be higher in the cooling cycle and decrease under the influence of the magnetic field. However, the change in $\rho_{ph}$ as a function of thermal cycle and the magnetic field are much more pronounced. The variation of the coefficient $\rho_{ph}$ with the H in both the thermal cycles is plotted in Fig. 10. A sharp decrease in the value of $\rho_{ph}$ in warming cycle as well as with magnetic field in both cooling and warming cycle shows the weakening of the electron-lattice coupling during the warming cycle as well as under magnetic field. This leads to substantially lower resistivity and considerably higher $T_C/T_{IM}$ in the warming cycle as well as at higher values of H.

**Conclusions**

We deposited LPCMO (y=0.4) epitaxial thin film on STO by RF-magnetron sputtering. HRXRD reveals that the films are epitaxial but contain high density of lattice defects such as dislocation networks which are responsible for relaxation of substrate induced strain. These observations are also supported by AFM measurements. The competition between co-existing FMM and AFM/COI phases governs the temperature dependent magnetic and transport properties of LPCMO. The protocol (ZFC, FCC and FCW) dependent FM transitions, strong divergence between the ZFC and FCW M (T) and prominent irreversibility between the FCC-FCW M (T) provide evidence in favor of the non-ergodic behavior of the magnetic state in this film. This nonergodicity, the origin of which could be traced to the delicate coexistence of FMM and AFM/COI phases leads to the existence of a liquid like behaviour. This liquid like state gets frozen into SRG. M (H) measurements also confirmed the frozen glassy states at low temperatures and unblocking of these frozen states with increasing temperature. The $\rho$-T measurements also confirm the liquid like behaviour and frozen glassy state at low temperatures. The liquid like behaviour is also confirmed by the large drop in resistivity under moderate magnetic fields. The small polaron activation energy decreases nonlinearly with increase in H and the variation of $T_{IM}$ with the activation energy remains nearly linear in the warming cycle, but it becomes nonlinear in the cooling cycle. The analysis of the low temperature $\rho$-T data reveals the weakening of the electron-lattice coupling under the influence of an external magnetic field as well as during the warming cycle. The devitrification of the frozen glassy state gives rise to Kondo like resistivity minimum.

Acknowledgements

Authors are grateful to Prof. R. C. Budhani for his persistent encouragement. Dr. Anurag Gupta and Dr. V. P. S. Awana are thankfully acknowledged for magnetic (MPMS-DST facility) and magnetotransport measurement, respectively.

**Figure captions**

Figure 1.        Experimental and simulated X-ray reflectivity (XRR) plot of LPCMO film.

Figure 2.        X-ray diffraction pattern of LPCMO film on STO. Inset shows φ-scan of (001) plane of LPCMO film and STO substrate.

Figure 3.        Rocking curve (ω -scan) along (110) and (001) planes of LPCMO film.

Figure 4.        Tapping mode AFM image of annealed LPCMO film. Inset shows AFM image of as grown film.

Figure 5.        Plot of magnetization as a function of applied magnetic field (M (H) curves) of LPCMO film at 10 K and 50 K.

Figure 6.        Resistivity as a function of temperature (ρ (T) curves) of epitaxial LPCMO film measured at magnetic field (H) of 0, 10, 30 and 50 kOe.

Figure 7.        Variation TIM as function of applied magnetic field (H) in cooling and warming cycles. Inset shows difference if TIM as a function of H.

Figure 8        (a) Plot of variation of activation energy ($E_A$) as a function of applied magnetic field (H) and (b) Plot of variation of $T_{IM}^C/T_{IM}^W$ with $E_A$.

Figure 9        The low temperature ρ-T data measured in cooling and warming cycle at H = 0 kOe and H = 30 kOe. The open symbols are the experimental data and the solid line are the fitted data.\

Figure 10.      Variation of the fitting parameter $\rho_{ph}$ measured in cooling and warming cycle as a function of the applied magnetic field.

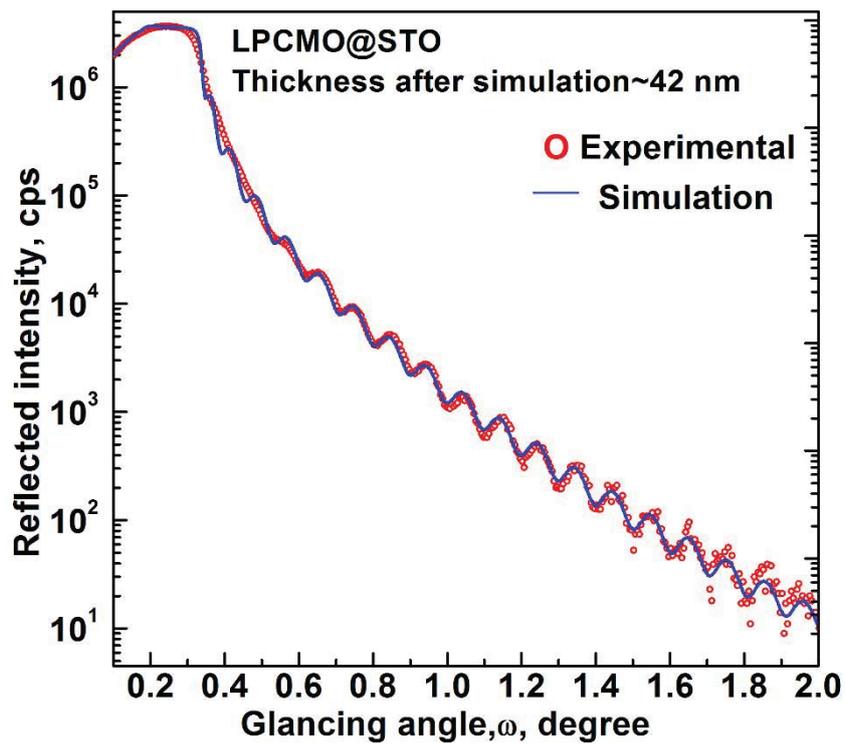

Fig. 1

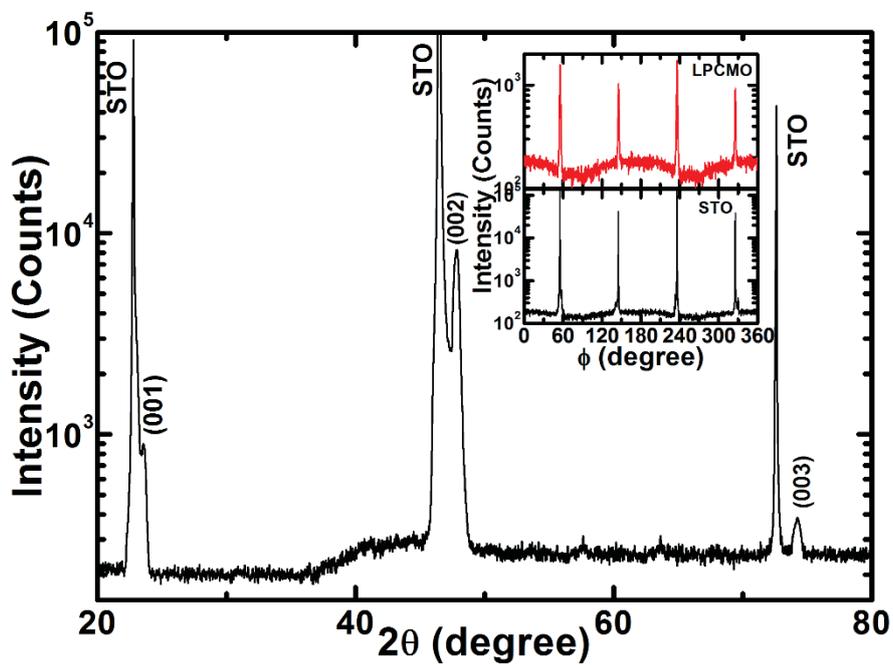

Fig. 2

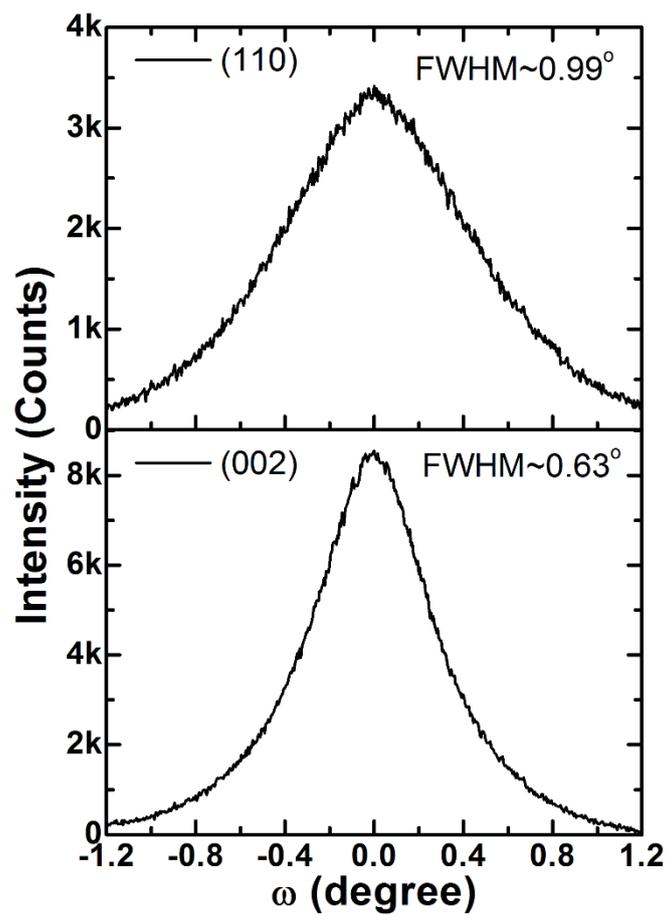

Fig. 3

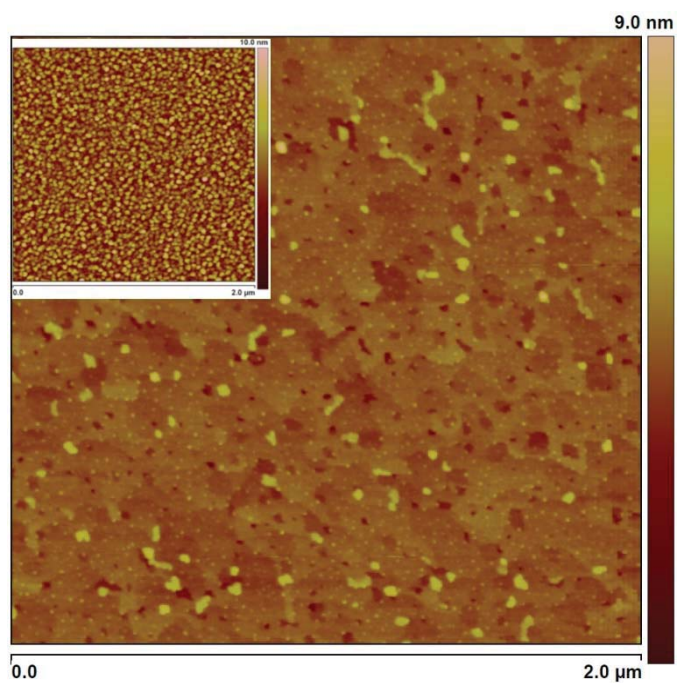

Fig. 4

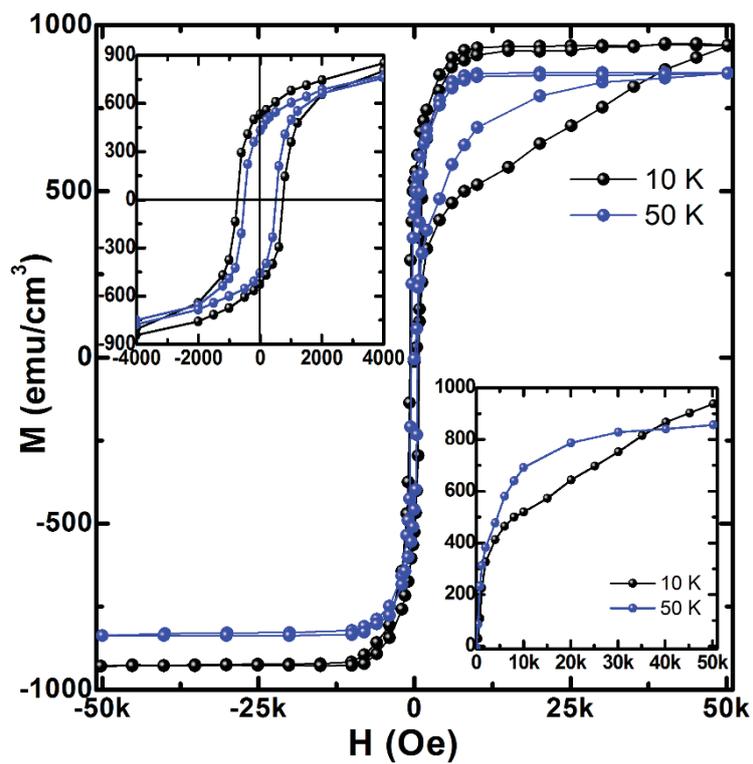

Fig. 5

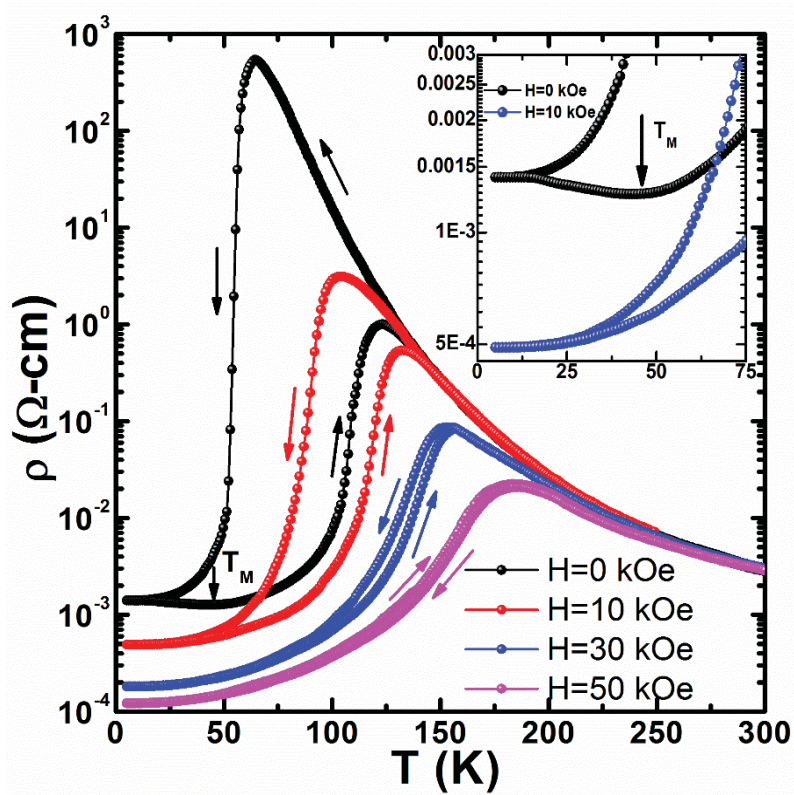

Fig. 6

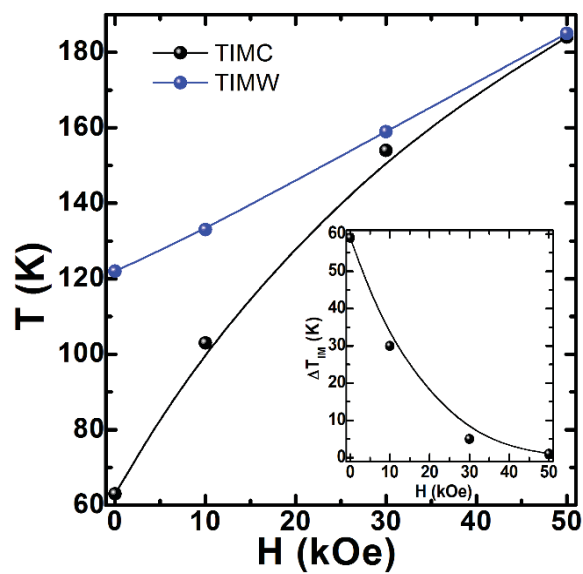

Fig. 7

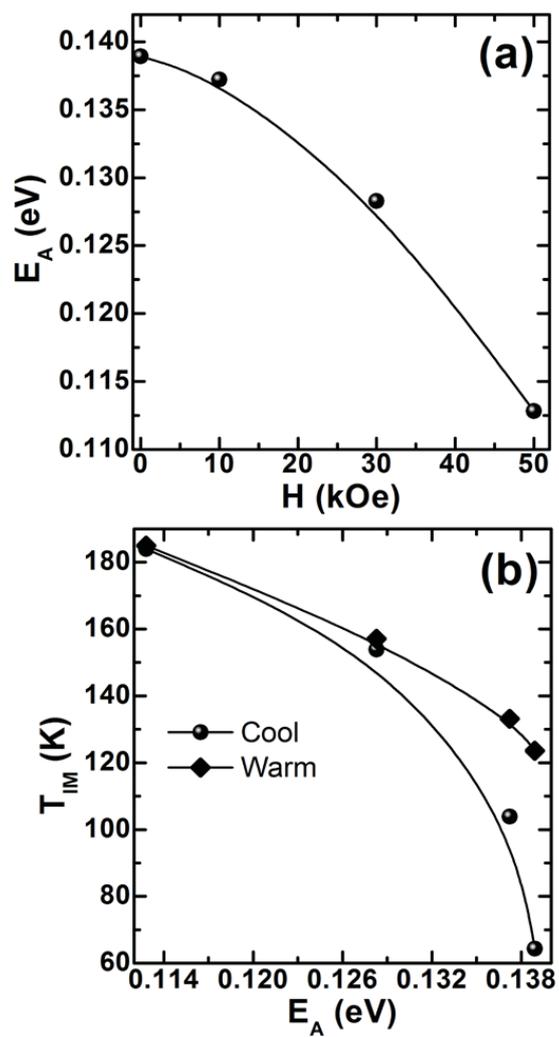

Fig. 8

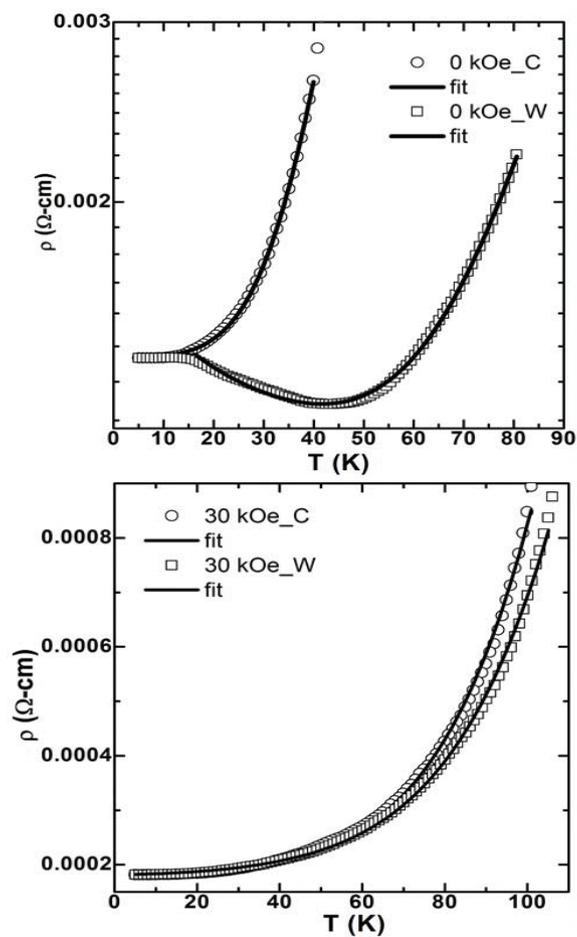

Fig. 9

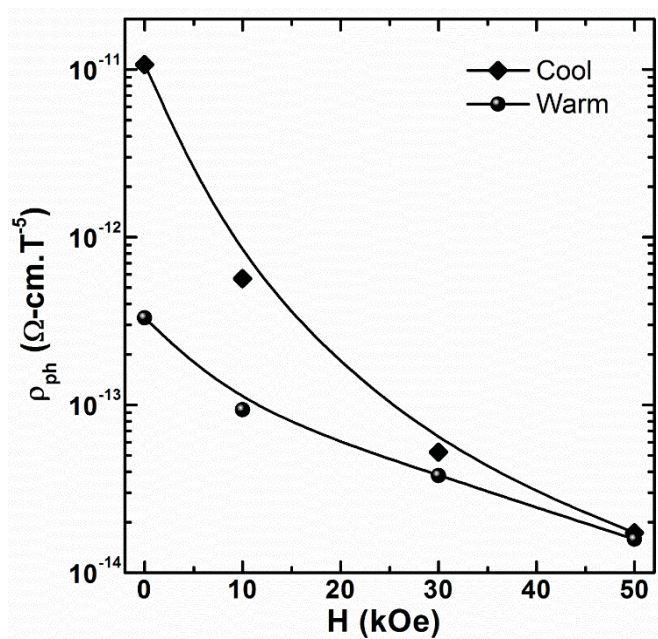

Fig. 10